\documentclass[twocolumn,showpacs,preprintnumbers,amsmath,amssymb]{revtex4}

\usepackage{graphicx}
\usepackage{dcolumn}
\usepackage{bm}

\begin{document}


\title{Improved setup for producing slow beams of cold molecules using a rotating nozzle}

\author{M. Strebel}
\author{F. Stienkemeier}
\author{M. Mudrich}
\email{Marcel.Mudrich@physik.uni-freiburg.de}
\affiliation{Physikalisches Institut, Universit\"at Freiburg, 79104 Freiburg, Germany}

\date{\today}

\begin{abstract}
Intense beams of cold and slow molecules are produced by supersonic expansion out of a rapidly rotating nozzle, as first demonstrated by Gupta and Herschbach~\cite{Gupta:1999, Gupta:2001}. An improved setup is presented that allows to accelerate or decelerate cold atomic and molecular beams by up to 500\,m/s. Technical improvements are discussed and beam parameters are characterized by detailed analysis of time of flight density distributions. The possibility of combining this beam source with electrostatic fields for guiding polar molecules is demonstrated.
\end{abstract}

\pacs{Valid PACS appear here}
\maketitle

\section{\label{sec:Intro}Introduction}
Cold molecules currently attract great attention due to their potential application for precision measurements~\cite{Hudson,Krems:2009}, quantum information schemes~\cite{DeMille,Krems:2009}, degenerate quantum gases with complex interactions~\cite{Baranov,Krems:2009}, and for cold chemistry~\cite{Weck:2006,Bodo:2004,Willitsch:2008,Smith:2008,Krems:2005,Krems:2009}. Besides the approach of forming cold molecules out of a sample of ultracold trapped atoms~\cite{Masnou:2001,Jones:2006,Chin:2009,Krems:2009}, recent experimental efforts mainly focus on developing techniques of slowing or filtering, cooling, and trapping molecules produced in effusive or nozzle beams. Well established techniques are buffer gas cooling \cite{Doyle:1995,Krems:2009}, the deceleration of beams of polar molecules using time-varying electric fields~\cite{Bethlem1:1999,Hutson:2006,Krems:2009}, as well as filtering of polar molecules out of an effusive source using static or time-varying electric fields~\cite{Rangwala:2003,Junglen:2004}. Recently, the method of decelerating supersonic molecular beams has been extended to magnetic~\cite{VanHaecke:2007,Narevicius:2008} and optical fields~\cite{Fulton:2006}. Alternative routes to producing cold molecules have been demonstrated utilizing the kinematics in elastically or reactively colliding molecular beams~\cite{Elioff:2003,liu:2007,Smith:2008}.

Furthermore, producing beams of slow and cold atoms and molecules by mechanical means has been demonstrated~\cite{Gupta:1999,Gupta:2001,Narevicius:2007}. In particular the technique of translating a supersonic jet to low longitudinal velocities by means of a rapidly counter-rotating nozzle, pioneered by Gupta and Herschbach, holds the promise of producing cold, slow and intense beams of nearly any molecular species available as a gas at ambient temperatures~\cite{Gupta:1999, Gupta:2001}. Attempts to improve the original arrangement have been made by the groups of H.-J. Loesch in Bielefeld and of M. DeKieviet in Heidelberg~\cite{LangThesis:2002}. By controlling the sense and speed of rotation of the nozzle as well as by using the seeded beams technique, tunable beam velocities ranging from thousands down to a few tens of meters per second are feasible. Typically, the translational and internal temperatures in the moving frame of the slowed beam are in the range of a few Kelvin and a few tens of Kelvin, respectively. Due to the extremely simple concept, this technique can be realized by a relatively simple mechanical device. Therefore this versatile source of cold molecules may be particularly useful for novel studies of molecular reaction dynamics at low collision energies in the gas phase or with surfaces~\cite{Weck:2005,Weck:2006,Bodo:2004,Liu:2004}.

The principle of operation relies on the Galilei transformation of particle velocities in the moving frame of the rotating nozzle into the resting laboratory frame. As the nozzle moves backwards with respect to the molecular beam motion the beam velocity $v_s$ is reduced to $v_0=v_s-v_{rot}$. In this way, Gupta and Herschbach succeeded in slowing beams of krypton (Kr) and xenon down to 42 and 59 m/s, respectively, at beam temperatures down to 1.5 and 2.6 K, respectively~\cite{Gupta:2001}. In their prototype setup they used a tapered aluminium barrel with a hollow bore as a rotor. The special aluminum alloy was chosen for its favorable ratio of tensile strength to mass density, allowing high rotational speeds. However, machining such a rotor barrel requires considerable efforts. The gas injection was realized by inserting a polymer needle into the rotor. However, erosion of this type of sealing and the resulting gas leakage was found to be a serious limitation. The nozzle orifice was fabricated by gluing a stainless steel disk with a 0.1\,mm aperture close to the tip of the rotor. In this case, swatting of slowly emerging molecules at high rotational speeds by the rotor arm as it comes back after one full revolution may be an issue. Furthermore, a general drawback of producing slow beams using a moving source is the fact that the flux of molecules $\dot{N}_0$ on the beam axis drops down sharply with decreasing laboratory beam velocity $v_0$ and with increasing distance $d$ of the detector from the nozzle, $F(v_{0}, d)\propto (v_0/d)^2$. This is a consequence of the beam spreading in transverse directions which becomes more and more important as $v_0$ decreases and hence the flight time goes up. The aim of the improved setup presented in this paper is to remedy the drawbacks mentioned above.

\section{\label{sec:Setup}Experimental setup}
\begin{figure}
\includegraphics[width=6cm]{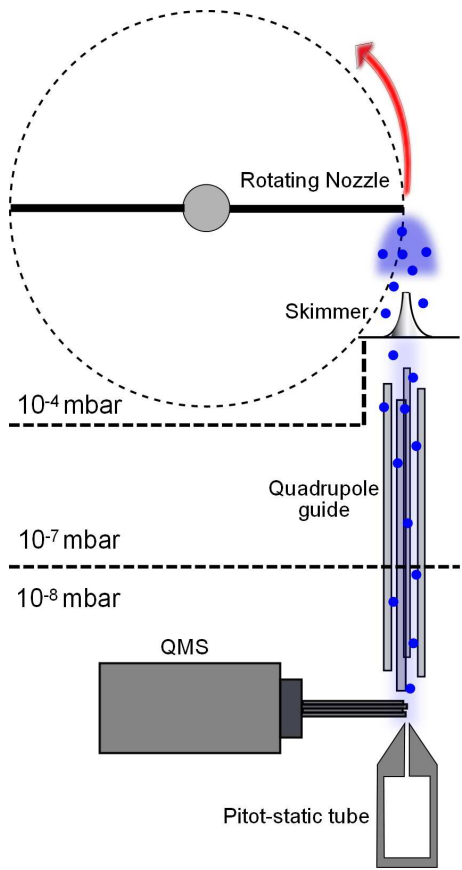}
\caption{\label{fig:schematic} Schematic overview over the experimental arrangement used for characterizing the rotating nozzle molecular beam source.}
\end{figure}
The experimental arrangement used for characterizing the new rotating nozzle setup is schematically depicted in Fig.~\ref{fig:schematic}. The upper part represents the moving molecular beam source including a hollow rotor and a skimmer with 1\,mm diameter. The nozzle chamber is pumped by an 8000\,l/s oil diffusion pump backed by roots and rotary vane pumps. Typical background gas pressure during operation with argon (Ar) at $p_0=0.3$\,bar nozzle pressure is $p_{bg}\approx 10^{-4}$\,mbar. The molecular beam follows a trajectory through the following elements: At a distance of about 20\,mm behind the nozzle it enters an intermediate chamber through a skimmer with 1\,mm opening diameter. The intermediate chamber is pumped by a 600\,l/s turbo pump providing a background gas pressure $p_{detect}\approx 5\times 10^{-9}$\,mbar during operation. The intermediate chamber is connected to a detector chamber pumped by a 150\,l/s turbo pump to maintain a background pressure $p_{detect}\lesssim 10^{-8}$\,mbar.

For the purpose of guiding beams of polar molecules, a set of four parallel steel rods is installed around the beam axis between the skimmer and a commercial quadrupole mass spectrometer (QMS) (Pfeiffer QMS200) which we use as a detector (see Sec.~\ref{sec:Guide}). The mounting discs used for holding the rods also serve as apertures for differential pumping between the intermediate chamber and the detector chamber. The total fight distance from the nozzle to the detector amounts to $\ell =35$\,cm. Since the positions of the quadrupole guides and the QMS detector cannot be adjusted, the beam axis is initially aligned by sight by adjusting the position of the skimmer. For additional characterization of the beam, a Pitot tube is installed about $5\,$mm behind the ionizer of the QMS detector. It consists of a cylindrical chamber 4\,cm in diameter and 10\,cm in length having a conical aperture on the side close to the QMS and a cold cathode ion gauge on the opposite side as an end cap. When a continuous molecular beam (nozzle at rest) is directed into the detector chamber, the residual gas pressure inside the Pitot tube rises until equilibrium between incoming flux and outgoing flux of molecules effusing out of the Pitot tube aperture according to its molecular flow conductance is reached. Thus, from the difference between pressures inside the Pitot tube and in the QMS chamber, $\Delta p$, the flux of molecules on the beam axis, $\dot{N}_0$, can be inferred according to $\dot{N}_0=\Delta p A a \bar{v}/(4k_{\mathrm{B}}T)$. Here $A$, $a$, and $\bar{v}$ denote the area of the entrance channel into the Pitot tube, a scaling parameter to account for the channel length, and the mean velocity of the molecules at ambient temperature $T$.
The on-axis beam density, which is obtained from the molecule flux using $n=\dot{N}_0/(v_0 A)$, is compared to the one determined by the QMS signals in Sec.~\ref{sec:Results}.

\begin{figure}
\includegraphics[width=7cm]{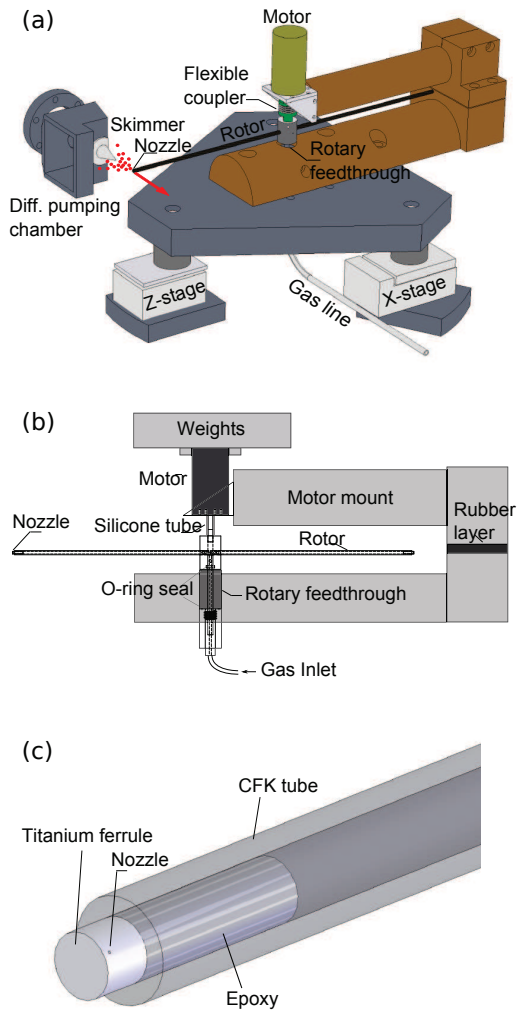}
\caption{\label{fig:setup} Overview (a) and close-up views (b,c) of the experimental setup. (a) The nozzle orifice is mounted on one tip of the hollow rotor and produces molecule pulses along the beam axis each time it passes nearby the skimmer. The gas is injected into the rotor through a rotary feedthrough from below. The motor is mounted to a massive brass mount on top of the rotor, to which it is connected by a flexible coupler (silicone tube). (b) Sectional view of the central part of the setup. (c) Close-up view of the rotor tip, which contains a bored thin-walled titanium ferrule as a nozzle.}
\end{figure}
The experimental realization of our rotating source of cold atoms and molecules is depicted in Fig.~\ref{fig:setup} (a). The following main modifications with respect to the work of Gupta and Herschbach have been implemented: The rotor is made of a thin carbon fiber (CFK) tube with 4 mm outer diameter and 38 cm in length. CFK is superior to any metallic material regarding the ratio of tensile strength to density by at least a factor of $4$. Moreover, the small mass of the CFK tube makes it simpler to suppress deterioral mechanical vibrations that may occur due to slight imbalance of the rotor at high speeds of rotation. The CFK tube is sealed on both ends by gluing thin-walled (0.25\,mm) titanium grade 5 ferrules as end-caps into the inner surface of the CFK tube using special epoxy for CFK (Fig.~\ref{fig:setup} (c)). One of the ferrules has a borehole $d=0.1\,$mm in diameter at a distance of 0.5\,mm from the closed end, thus forming a well defined nozzle aperture. A simple alternative that we also tried is capping the CFK tube by gluing circular pieces of capton foil to front and back sides of the tube. Capping the CFK tube this way has proven to be stable enough to withstand the high centrifugal forces acting at rotational speeds up to $f_{rot}=350$\,Hz (21000 rpm). The nozzle orifice was fabricated by inserting a $50\,\mu$m wire half way in between the CFK front side and the capton foil prior to gluing. After the epoxy has hardened the wire is pulled out and an approximately 1\,mm long channel remains. This method allows to easily realize small nozzle channels of diameters $<100\,\mu$m, which limits the gas load and thus the required pumping speed. Moreover, using this method of fabrication the distance of the nozzle from the tip of the rotor is kept minimum, thereby suppressing swatting of slow molecules by the rotor. However, the expansion out of a long nozzle channel is unfavorable compared to a thin-walled orifice as far as adiabatic cooling conditions are concerned. Besides, nozzles fabricated using this procedure vary considerably with respect to length and shape of the nozzle channels. For all measurements presented in this work, bored thin-walled titanium ferrules are used.

The CFK tube is inserted and glued into the core of the rotor consisting of an aluminium or brass cylinder (Fig.~\ref{fig:setup} (b)). The latter has a borehole along its axis for inserting the bored shaft of a rotary feedthrough from below and the shaft of a flexible coupler attached to a motor (faulhaber-group.com) from above. In order to avoid any imbalance it is important to perfectly align the motor axis and the feedthrough axis. This is ensured by inserting the two axes into the same through-hole through the rotor core. The rotor including the metal core is first balanced statically by suspending the rotor core between two needles in the center of gravity and by abrading CFK material from the heavier side of the rotor until it is perfectly balanced. In a second step, the rotor is dynamically balanced by adding screws to the rotor core at different places and by monitoring the vibration spectrum using acceleration sensors attached to the setup. Using this procedure, vibrations at high rotation speeds resulting from the rotor are small compared to vibrations induced by the motor itself. Therefore, additional masses are attached to the motor (Fig.~\ref{fig:setup} (b)), which reduce vibrations at frequencies $\gtrsim 100$\,Hz.
Gas injection into the rotor is realized using a commercial rotary feedthrough sealed by ferrofluidic metal sealings. The bored shaft of the feedthrough is inserted into the rotor core and O-ring sealed. Metal-sealed feedthroughs (ferrotec.com) features very high leak tightness ($\lesssim 10^{-11}\,$mbar\,l/s) at low friction torque.

The rotor axis is aligned with the motor axis by connecting the rotary feedthrough and the motor to a U-shaped brass construction. This brass mount is broken in two parts to introduce a rubber layer in between the two parts as a damping element for suppressing vibrations. The brass mount is fixed to a triangular massive metal plate supported by neoprene dampers. In our new setup, the neoprene damper positioned underneath the skimmer resides on a vertical translation stage and the remaining two dampers reside on two horizontal translation stages. The translation stages are actuated by turning knobs outside of the vacuum chamber which allows full adjustment of the nozzle position with respect to the beam axis even during operation (cf. Fig.~\ref{fig:TwoPeaks}).

In our present setup the maximum speed of rotation is limited to about 350\,Hz (21000\,rpm). The limiting factor so far is mechanical vibrations, which rapidly increase in amplitude at frequencies $\gtrsim 300\,$Hz. The origin of these vibrations is presumably a slight unbalance of the rotor as well as vibrations caused by the motor. These vibrations are found to reduce the lifetime of the ferrofluidic sealings of the feedthrough, which is specified to rotation frequencies $\lesssim 166\,$Hz (10000\,rpm). The motor as well as the CFK rotor, however, are able to withstand much higher frequencies up to 970 and 800\,Hz, respectively.

\section{Analysis of time of flight data}
\label{sec:theory}
\begin{figure}
\includegraphics[width=8cm]{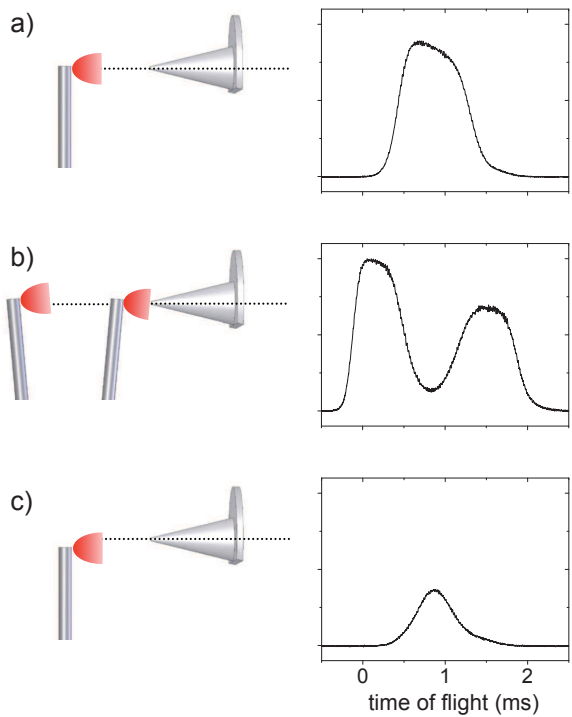}
\caption{\label{fig:TwoPeaks} Illustration of the dependence of measured signal (right side) on the alignment of the rotating nozzle setup with respect to the beam axis (dotted line). (a) Proper nozzle alignment; When the rotor is set to a perpendicular position with respect to the beam axis, nozzle and skimmer are in line with the beam axis up to the detector. (b) Nozzle shifted above the beam axis; Two pulses are produced each time the nozzle intersects the beam axis at two slightly tilted rotor positions. (c) Nozzle shifted below the beam axis; One pulse is measured with reduced amplitude and width.}
\end{figure}
Although the expansion out of the rotating nozzle is continuous, the beam passing through the skimmer and entering the detector chamber is chopped.
Each time the nozzle comes close to the skimmer part of the ejected particles travel along trajectories that match the beam axis, determined by the positions of skimmer, quadrupole guides, and QMS detector. Using the vertical and horizontal translation stages the rotor position can be aligned during rotation of the nozzle to optimize the detector signal. A peculiarity of the rotating-nozzle arrangement is the fact that the signal traces as a function of time are changed in an asymmetric manner as the rotor is shifted horizontally across to the beam axis, as illustrated in Fig.~\ref{fig:TwoPeaks}. The signal traces are recorded using Ar gas expanding out of the nozzle at a nozzle pressure $p_0=200$\,mbar and at a frequency of rotation of 10\,Hz. At this rotation frequency, the flight time of Ar atoms to the detector is short compared to the nozzle motion in front of the skimmer. Thus, the recorded signal traces mainly reflect the effective slit opening function given by the skimmer with respect to the moving nozzle. In Fig.~\ref{fig:TwoPeaks} (a), the nozzle is properly aligned such that the signal trace features a slightly asymmetric trapezoidal shape.
As the nozzle is moved across the beam axis by 2\,mm, the measured signal splits into two maxima, as shown in Fig.~\ref{fig:TwoPeaks} (b).
In this geometry, particle trajectories matching the beam axis occur at two different angles of inclination of the rotor with respect to the beam axis.
Misalignment by 2\,mm in the other direction leads to a reduction of the signal width and amplitude, as shown in Fig.~\ref{fig:TwoPeaks} (c). The inclination signal envelopes in Fig.~\ref{fig:TwoPeaks} (a) and (b) results from the decreasing solid angle defined by nozzle and skimmer aperture as the nozzle backs off away from the skimmer.

\begin{figure}
\includegraphics[width=9cm]{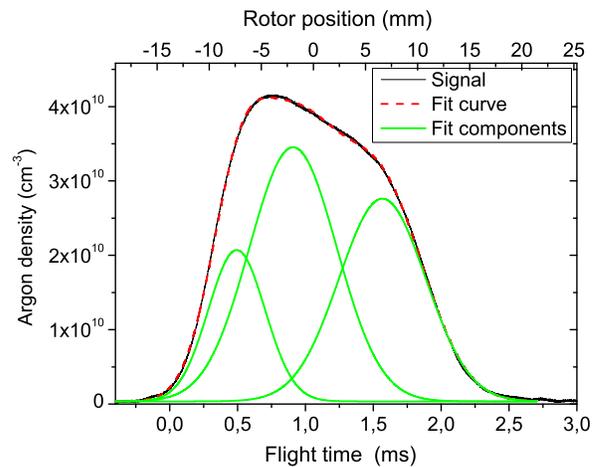}
\caption{\label{fig:Slit} Slit opening function measured at low rotation speeds of the nozzle (10\,Hz) (solid line). The dashed line represents a model curve fitted to the data, consisting of the sum of 3 Gaussian functions (light solid lines).}
\end{figure}
In order to extract the exact values of the beam velocity, temperature, and density from the measured signal traces, the effective slit opening function has to be characterized quantitatively and included into a model for the time-dependent density distribution of the molecular beam. The effective slit opening function for optimum horizontal nozzle alignment measured at low rotor frequency (10\,Hz) is plotted in Fig.~\ref{fig:Slit} (solid line).
At such low rotor frequency longitudinal dispersion of the beam pulse during flight time is negligible. This is verified
measuring the signal intensity at rest as a function of the rotor position using a micrometer driven manipulator. The resulting curve has identical shape compared to the time of flight curves at low rotor velocities. The dashed red line in Fig.~\ref{fig:Slit} represents a fitted model function consisting of the sum of 3 Gaussian functions,
\begin{equation}\label{eq:slit}
s_z(z)=\sum_{i=1}^{3} A_ie^{-\frac{(z-z_i)^2}{\Delta z_i^2}},
\end{equation}
where amplitude factors $A_i$, center positions $z_i$, and widths $\Delta z_i$ are determined by the fit to the experimental data. The density distribution in phase-space of a gas pulse produced by the rotating nozzle is modeled by the expression
\begin{equation}\label{phasespace}
f_0(\vec{r},\vec{v})=N\,f_r(\vec{r})\,f_v(\vec{v}),
\end{equation}
which is normalized to the total number of molecules per bunch, $N$.

The spatial distribution $f_r$ is assumed to be anisotropic according to
\begin{equation}\label{spatial}
f_r(\vec{r})=f_{\perp}(x,y)\,f_z(z).
\end{equation}
The distribution in transverse direction,
\begin{equation}\label{spatial}
f_{\perp}(x,y)=f_{\perp}^0\,e^{-\frac{x^2+y^2}{\Delta r^2}},
\end{equation}
is normalized to $f_{\perp}^0=(\pi\Delta r^2)^{-1}$. The distribution along the beam axis ($z$-axis) is obtained from the slit opening function (Eq.~(\ref{eq:slit})) by rescaling the z-coordinate according to $z\rightarrow z/q$, where $q=|v_s/v_{rot}|$ denotes the ratio between beam velocity $v_s=\sqrt{\kappa/(\kappa-1)}\hat{v}$ and the velocity of the rotor tip, $v_{rot}=2\pi R f_{rot}$. Here, $\kappa=5/3$ for monoatomic gases, $\hat{v}=\sqrt{2k_{\mathrm{B}}T_0/m}$ is the most probable velocity of the molecules inside the nozzle at a temperature $T_0$, and $R=19\,$cm stands for the rotor radius. Thus,
\begin{equation}\label{spatial}
f_z(z)=f_z^0/q\,\sum_{i=1}^3 A_ie^{-\frac{(z-z_i q)^2}{(\Delta z_iq)^2}},
\end{equation}
where the amplitude factor $f_z^0$ is given by $f_z^0=(\sqrt{\pi}\sum_{i=1}^3 A_i\Delta z_i)^{-1}$. The total number of molecules per pulse (integral over space of $f_r(\vec{r})$) is a function of speed of rotation,
$N=\dot{N}\Delta t$, where $\dot{N}=A\,p_0/(k_B T_0)\,\hat{v}\,\sqrt{2\kappa/(\kappa+1)}((\kappa+1)/2)^{1/(1-\kappa)}$~\cite{Pauly:2000} is the flux of molecules out of the nozzle orifice of area $A=\pi (d/2)^2$ at pressure $p_0$ and $\Delta t =\Delta z_{slit}/v_{rot}\approx 2\,$cm$/v_{rot}$.
In other words, as the nozzle velocity $v_{rot}$ falls below the beam velocity $v_s$, the molecule bunches become longer than the spatial width of the slit opening function $\Delta z_{slit}$ and the number of molecules per bunch increases accordingly.

The velocity distribution is chosen according to the most frequently used ellipsoidal drifting Maxwellian model $f_v(\vec{v})=f_{v_{\perp}}(v_x,v_y)\times f_{vz}(v_z)$, where
\begin{equation}\label{velocitydist}
f_{v_{\perp}}=f_{v_{\perp}}^0\,e^{-\frac{v_{\perp}^2}{\Delta v_{\perp}^2}},\,\,\mathrm{and}\,\,f_{vz}=f_{vz}^0\,e^{-\frac{(v_z-v_0)^2}{\Delta v_z^2}},
\end{equation}
with normalization constants $f_{v_{\perp}}^0=(\pi\Delta v_{\perp}^2)^{-1}$ and $f_{vz}^0=(\sqrt{\pi}\Delta v_z)^{-1}$~\cite{Scoles:1988, Pauly:2000}. The widths of the velocity distributions, $\Delta v_{\perp}$ and $\Delta v_{z}$ are related to the transverse and longitudinal beam temperatures $T_{\perp}$ and $T_{\|}$ by $T_{\|,\perp}=m \Delta v_{\|,\perp}^2 /(2 k_B)$. While $T_{\|}$ results from adiabatic cooling of the gas in the jet expansion, $T_{\perp}$ is an approximate measure of the transverse beam divergence. Typical values of $T_{\|}$ attainable in free jet expansions range between $T_{\|}\sim 1$\,K and $T_{\|}\sim 10$\,K, while the transverse beam divergence corresponds to $T_{\perp}\sim 300\,$K.
During free expansion, in the absence of collisions and external forces, the distribution function remains unchanged according to Liouville's theorem. The time evolution is given by
\begin{equation}\label{phasespacetime}
f(\vec{r},\vec{v},t)=f_0(\vec{r}-\vec{v}t,\vec{v}).
\end{equation}
The spatial density as a function of time is then obtained by integrating over velocities,
\begin{equation}\label{vintegral}
n(\vec{r},t)=\int d^3v f(\vec{r},\vec{v},t).
\end{equation}

The analytic solution of this expression is given by
\begin{equation}\label{eq:densitytime}
n(\vec{r},t)=N n_{\perp}(x,y) n_z(z),
\end{equation}
where
\begin{equation}
n_{\perp}=\pi\left(\frac{1}{\Delta v_{\perp}^2}+\frac{t^2}{\Delta r^2}\right)^{-1}\,e^{-\frac{x^2+y^2}{\Delta r^2+\Delta v_{\perp}^2t^2}}
\end{equation}
and
\begin{equation}\label{eq:zdensitytime}
n_{z}=\sqrt{\pi}\sum_{i=1}^3 A_i\left(\frac{1}{\Delta v_z^2}+\frac{t^2}{\Delta z_i^2}\right)^{-1/2}\,e^{-\frac{(z-z_i-v_0 t)^2}{\Delta z_i^2+\Delta v_z^2t^2}}.
\end{equation}

\begin{figure}
\includegraphics[width=8cm]{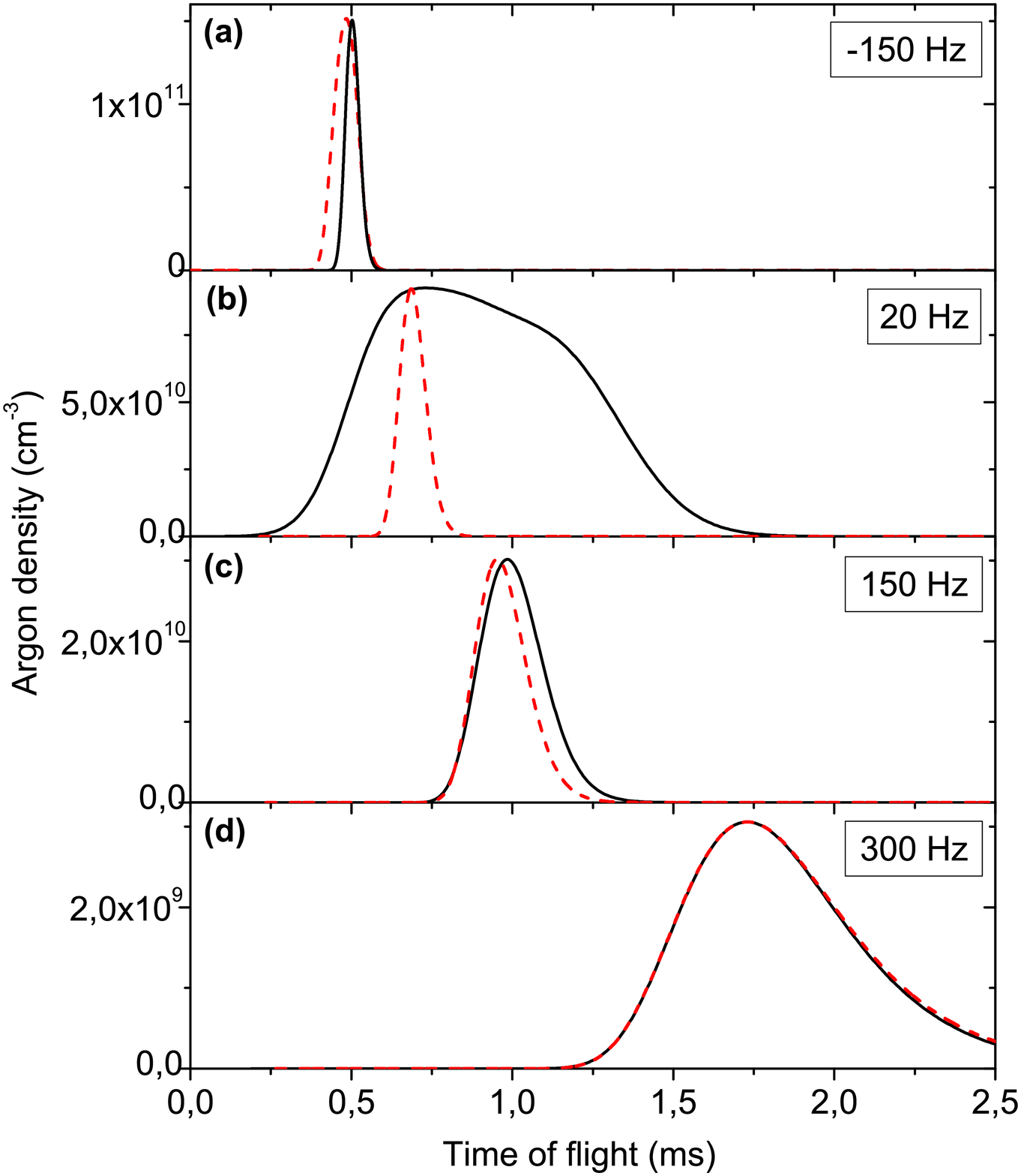}
\caption{\label{fig:Faltung} Simulated density distributions on the beam axis as a function of time of flight at various speeds of rotation of the nozzle. Solid lines are calculated according to Eq.~(\ref{eq:densitytime}) based on the measured slit-opening function, dashed lines reflect the longitudinal velocity distribution rescaled to flight times.}
\end{figure}
Fig.~\ref{fig:Faltung} displays simulated Ar density distributions on the beam axis, $n(0,0,\ell,t)$, according to Eq.~(\ref{eq:densitytime}) for typical beam temperatures $T_{\|}=5\,$K and $T_{\perp}=323\,$K at various speeds of rotation (solid lines). Here, $\ell =0.38\,$m denotes
the total flight distance from the nozzle to the QMS detector. The nozzle speed is $v_{rot}=-179$\,m/s ($-150\,$Hz) in Fig.~\ref{fig:Faltung} (a), 24\,m/s
(20\,Hz) (b), 179\,m/s (150\,Hz) (c), and 358\,m/s (300\,Hz) (d). Besides shifting linearly to longer flight times at higher rotational speeds,
the distribution is first broadened at rotations around 0\,Hz due to the effect of the slit opening function. At higher speeds, the distribution first narrows down
as the nozzle moves faster past the skimmer ($150\,$Hz), and then becomes broader again
as a consequence of longitudinal dispersion, \textit{i.\,e.} due to the fact that longer flight times allow the molecule bunches to longitudinally
broaden due to the finite beam temperature $T_{\|}$. The dashed lines represent the neat velocity distributions horizontally rescaled to flight times
according to $t=\ell /v_z$. Thus, at negative speeds of rotation (molecular beam is accelerated) the measured signal is predominantly determined
by the shape of the velocity distribution, while at speeds around 0\,Hz the measured trace is strongly distorted by the slit-opening function. At higher rotation frequencies,
the effect of the slit function diminishes and the two distributions eventually converge. Therefore, at rotations $f_{rot}\gtrsim 200\,$Hz, Eq.~(\ref{eq:zdensitytime})
may be replaced by the simpler expression $n'(t)=A'/t^2\,\exp\left( -(s/t-v_0)^2/\Delta v_z^2\right)$ in the procedure of fitting the measured data.

\section{\label{sec:Results}Beam parameters}
\begin{figure}
\includegraphics[width=8cm]{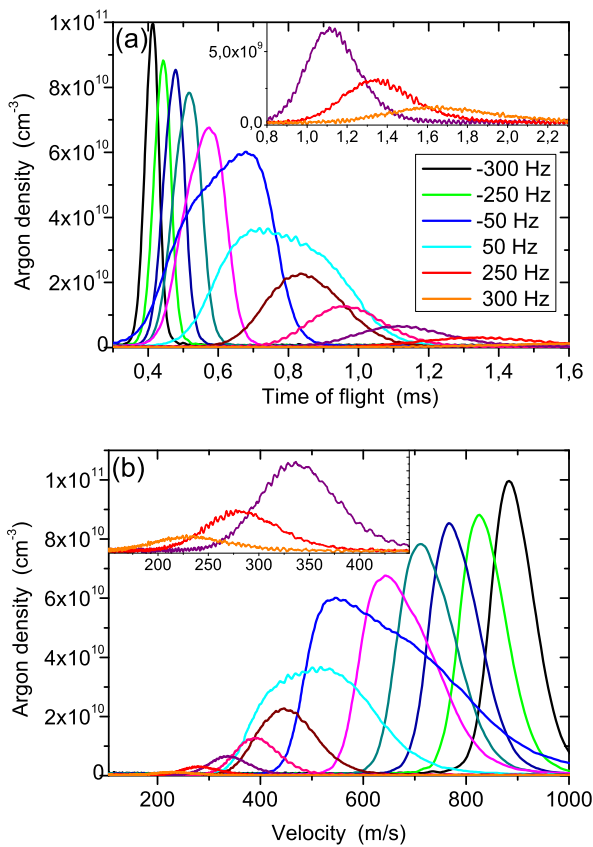}
\caption{\label{fig:TOFs}(a) Measured time of flight density distributions of decelerated argon beams at various speeds of rotation of the nozzle.
(b) Same data as in (a), where the horizontal axis is rescaled to velocities.}
\end{figure}
Typical time of flight distributions measured with Ar at a nozzle pressure $p_0=400$\,mbar are displayed in Fig.~\ref{fig:TOFs} (a) for various speeds
of rotation. In these measurements, the Ar gas pressure inside the gas line was carefully adjusted such that the effective pressure at the nozzle remains
constant. Due to the centrifugal force the nozzle pressure $p_0$ rises sharply at increasing speed of rotation $v_{rot}$ while holding the gas-line pressure $p_{0,gas\,line}$ constant according to
$p_{0}=p_{0,gas\,line}\exp\left( m v_{rot}^2/(2 k_B T_0)\right)$~\cite{Gupta:1999, Gupta:2001}. Signal broadening due to the convolution with the slit opening function around 0\,Hz as well as broadening due to longitudinal dispersion at high rotation frequencies
are clearly visible (\textit{cf.} inset of Fig.~\ref{fig:TOFs} (a)).
The time of flight distributions
shown in Fig.~\ref{fig:TOFs} (a) are horizontally rescaled to velocity distributions according to $v=\ell /t$ and replotted in Fig.~\ref{fig:TOFs} (b). In this
representation of the data the signal broadening due to dispersion at high rotation frequencies cancels out. However, broadening due to the slit opening function still remains and masks the real velocity distribution. The absolute densities are estimated by scaling the QMS detector signal to the output of the pressure gauge in the detector chamber.
Besides, densities can be estimated from the beam intensity measurement using the Pitot as discussed in Sec.~\ref{sec:Setup}. However, the density values obtained in this way turn out to be about a factor of 10 higher than the ones from the QMS measurement. This may be due to some feedback effect of the cold cathode ion gauge onto the vacuum inside the small volume of the Pitot tube, thus causing a nonlinear response of the vacuum gauge.

\begin{figure}
\includegraphics[width=8cm]{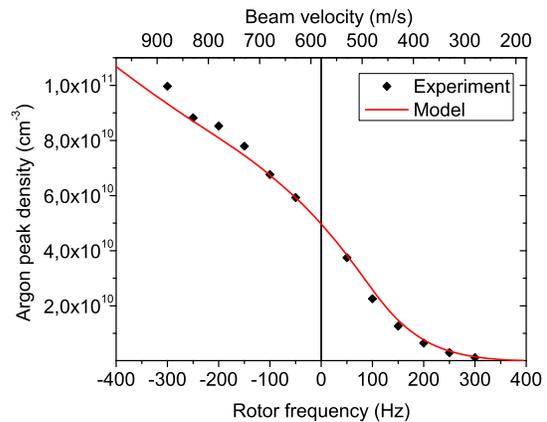}
\caption{\label{fig:PeakDens}Peak values of the time of flight argon density distributions as a function of the rotation frequency. Symbols are the measured values, the solid line represents the result of fitting (Eq.~(\ref{eq:densitytime})) to the data.}
\end{figure}
Strikingly, the Ar signal amplitude rapidly drops by about a factor of 30 when increasing the speed of rotation from 0\,Hz to 300\,Hz. In contrast, when the beam is accelerated by spinning the nozzle in the opposite direction up to 250\,Hz, the signal only slightly increases by a factor of 2. The dependence of
the measured peak density from the speed of rotation is shown in Fig.~\ref{fig:PeakDens} as filled symbols. The solid line represents maximum values of the model function, $\max\left(n\right)$,
according to Eq.~(\ref{eq:densitytime}). Here, the amplitude factor of the model function, $N$, is scaled by a factor of 0.5 to fit the experimental data. This strong amplitude modulation as a function of sense and speed of rotation is caused by both longitudinal as well as transverse beam dispersion.
Since the rate of the signal drop is determined by the beam temperatures $T_{\|}$ and $T_{\perp}$, it is evident that beam temperatures have to be kept
low by ensuring optimum free jet expansion conditions, \textit{i.\,e.} high effective nozzle pressure $p_{0,nozz}\gtrsim 100\,$mbar at sufficiently low
background pressure $p_{bg}\lesssim 10^{-3}\,$mbar as well as a high quality nozzle aperture and skimmer~\cite{Scoles:1988}.

\begin{figure}
\includegraphics[width=9cm]{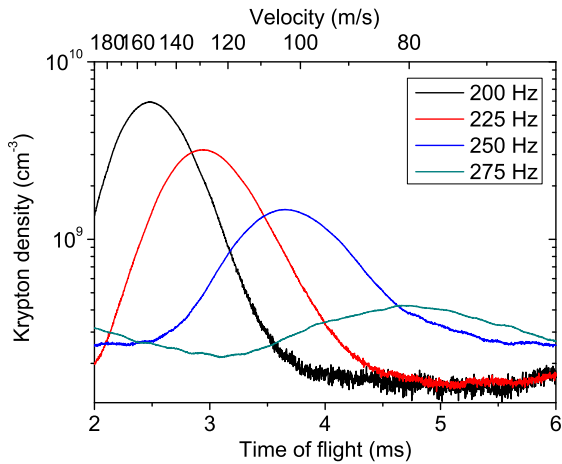}
\caption{\label{fig:KrTOF}Measured time of flight density distributions of krypton at various speeds of rotation of the nozzle.}
\end{figure}
Since the terminal beam velocity of jet beams is reduced with increasing mass of the beam particles, deceleration of beams of the heavier nobel gas atoms
krypton (Kr) and xenon to small velocities is achieved at lower
speeds of rotation. Beams of Kr, \textit{e.\,g.}, have terminal velocities around $400$\,m/s, such that deceleration down to velocities below 100\,m/s are easily
achieved at rotation frequencies $\gtrsim 250$\,Hz, as shown in Fig.~\ref{fig:KrTOF}. However, due to longitudinal dispersion, consecutive beam packets start to overlap
at frequencies $\gtrsim 250$\,Hz ("wrap around effect"), which impedes quantitative analysis of the time of flight distributions. One way of passing around
this effect could be the implementation of an additional mechanical beam chopper to pick out individual beam packets.

In comparison with other methods of producing beams of slow molecules, our results are quite similar regarding beam densities at low velocities. Using electrostatic quadrupole filters, \textit{e.\,g.}, typical peak densities around $10^9$\,cm$^{-3}$ at velocities around 60\,m/s are achieved~\cite{Motsch:2009}. However, the rotating nozzle setup offers the possibility of tuning beam velocities in a wide range. Due to free jet expansion conditions, the internal degrees of freedom of the molecules are expected to be cooled to the Kelvin range, which can be reached using quadrupole filters only with substantial experimental efforts~\cite{buuren:2009}. Besides, the rotating nozzle is suitable for any atomic of molecular substance available in the gas-phase at ambient temperatures, in contrast to filtering or deceleration techniques, which rely on particularly suitable Stark of Zeeman effects of molecules in external electrostatic or magnetic fields.

\begin{figure}
\includegraphics[width=8cm]{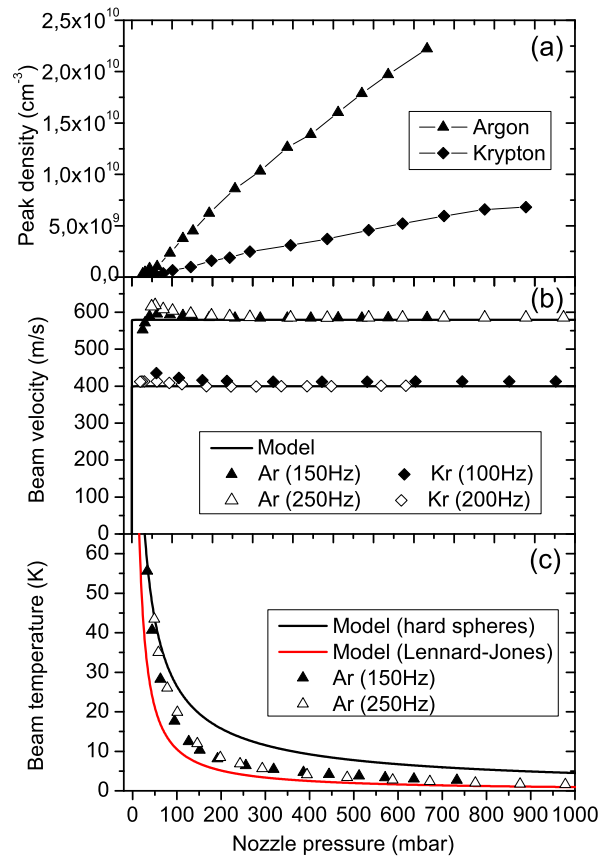}
\caption{\label{fig:TempVel} Argon and krypton beam parameters as a function of nozzle pressure.}
\end{figure}
In order to characterize the beam parameters and to test the analytic model, flight distributions of decelerated Ar and Kr beams are measured at various expansion pressures and speeds of rotation and are fitted by the model function, Eq.~(\ref{eq:densitytime}). Free fit parameters are the total number of atoms per bunch, $N$, the reduced beam velocity $v_0$, and the longitudinal beam temperature $T_{\|}$. The resulting values for Ar and Kr beams are depicted in Fig.~\ref{fig:TempVel} as symbols. Clearly, the results do not depend on the speed of rotation of the nozzle within the experimental uncertainty. As expected, the peak densities of Ar and Kr beams measured at rotation frequencies 150 and 100\,Hz, respectively, evolve roughly proportionally to the nozzle pressure $p_0$ (Fig.~\ref{fig:TempVel} (a)). The Kr densities fall behind the ones reached with Ar by about a factor of 3 due to the smaller flux of Kr out of the nozzle and due to the smaller longitudinal beam velocity of Kr. The fitted beam velocities $v_s$ shown in Fig.~\ref{fig:TempVel} (b) are found to perfectly match the theoretical values for supersonic beams, $v_s$, provided the nozzle pressure exceeds about 100\,mbar. The latter are given by $v_s=\sqrt{\frac{2\kappa}{\kappa -1}\frac{k_B T_0}{m}}\left(1-\left(\frac{p_{bg}}{p_0}\right)^{\frac{\kappa -1}{\kappa}}\right)$, where $T_0$ and $p_0$ are the nozzle temperature and pressure, $p_{bg}$ is the background gas pressure and $\kappa=c_P/c_V=1+2/f$ is the adiabatic exponent for a gas of particles with $f$ degrees of freedom~\cite{Pauly:2000}. For mono-atomic gas ($\kappa = 5/3$) and low background pressure $p_{bg}\ll p_0$ we obtain the simpler formula $v_s=\sqrt{ 5\,k_B T_0/m}$, which yields $v_s=579\,$m/s for Ar and $v_s=399$\,m/s for Kr at $T_0=323\,$K. At lower nozzle pressure $p_0$, the fitted values of $v_0$ slightly deviate from the model curve, indicating expansion conditions at the transition from the effusive to the free jet expansion regime.

Fit values of the longitudinal temperature $T_{\|}$ of Ar beams are shown in Fig.~\ref{fig:TempVel} (c) as symbols. The solid lines represent model curves according to two different collision models. Assuming hard-spheres (HS) collisions, the terminal speed ratio $S=v_s/\Delta v$ is given by $S=0.289(p_0 d \sigma/T_0)^{0.4}$, while atom-atom interactions according to the Lennard-Jones (LJ) potential yield $S=0.397(p_0d\epsilon^{1/3}r_m^2/T_0^{4/3})^{0.53}$ \cite{Pauly:2000}. Here, $\sigma=46.3$\,[$50.3$]\,$\mathrm{\AA}^2$ represents the HS-cross section, $\epsilon =141.6\,$[$50.3$]\,K$\times k_{\mathrm{B}}$ and $r_m = 3.76\,$[$4.01$]\,$\mathrm{\AA}$ stand for the potential well depth and the equilibrium distance of the LJ-potential for argon [Kr], respectively. In comparison with the experimental data, the model assuming interactions according to the LJ-potential appears to yield more precise values than the one assuming HS-collisions. However, the LJ-model systematically slightly under-estimates longitudinal temperatures, in contrast to the HS-model, which gives too large values. The measured temperature continuously decreases as a nozzle pressure up to about 1\,bar is applied, reaching values $T_{\|}\lesssim 1.5\,$K with both Ar and Kr. Evidently, high nozzle pressures are desirable if low temperatures are to be achieved.

\section{\label{sec:Problems}Detrimental effects}
\begin{figure}
\includegraphics[width=8cm]{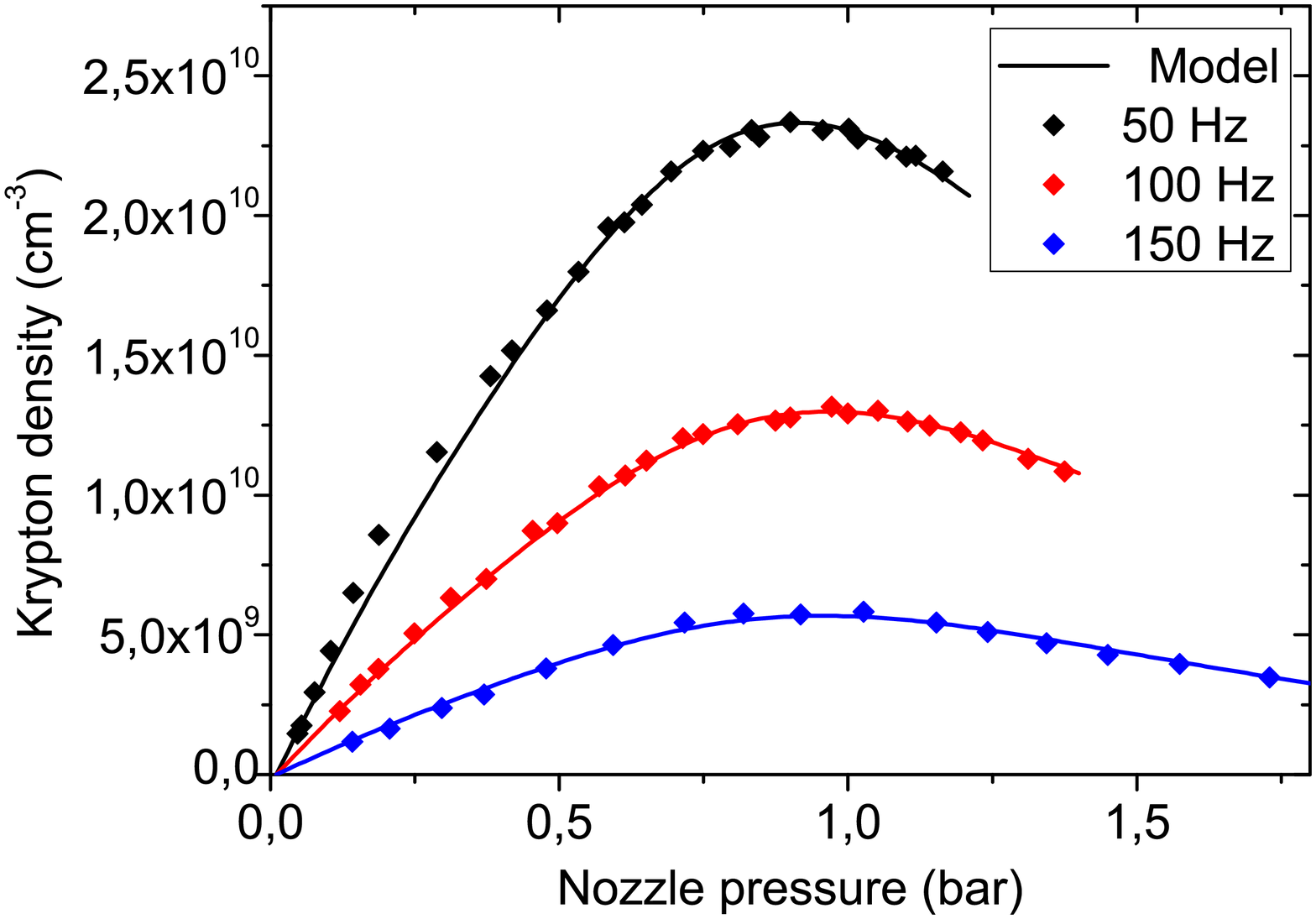}
\caption{\label{fig:KrAtt}Attenuation of the krypton peak beam intensity as a function of nozzle pressure for different speeds of rotation of the nozzle.}
\end{figure}
In order to work at high nozzle pressures while keeping the gas load low it may be favorable to use smaller apertures, which goes at the expense of beam intensity, though. The maximum applicable gas pressure, however, is limited by the pumping speed available in the rotor chamber. In our setup, the rotor-chamber pressure increases roughly linearly from $1.9\times 10^{-5}$\,mbar to $2.2\times 10^{-4}$\,mbar, as the Ar pressure at the nozzle rises from 100\,mbar up to 1000\,mbar. At these vacuum conditions, collisions of beam molecules with the ones from the residual gas may seriously deplete the beam. This effect is studied by measuring the peak beam density of Kr beams as a function of nozzle pressure up to $p_0\lesssim 2\,$bar at different rotation frequencies, $f=50\,$Hz ($v_0=340\,$m/s), $f=100\,$Hz ($v_0=280\,$m/s), $f=150\,$Hz ($v_0=220\,$m/s).

As shown in Fig.~\ref{fig:KrAtt}, the Kr beam density is affected by rest gas collisions as the nozzle pressure exceeds about 600\,mbar. Nozzle pressures higher than about 900\,mbar even lead to the destruction of the beam and beam densities drop down again, irrespective of the rotor frequency. Only at high frequencies leading to beam velocities below about $100$\,m/s the position of the maximum of the dependence $n(p_0)$ shifts down to lower values of $p_0$ due to the fact that slow molecules spend more time inside the region of high rest-gas pressure and are therefore more sensitive to rest-gas collisions. The solid lines in Fig.~\ref{fig:KrAtt} are fit curves according to the model of the peak density (Eq.~(\ref{eq:densitytime})), taking into account a density reduction due to collisions, $n_{red}=n\exp\left( -\sigma n_{bg}v_{rel}\,s/v_0\right)$. Here, the relative velocity between colliding atoms is taken as $v_{rel}=\sqrt{v_0^2+v_{bg}^2}$, $s\approx 2\,$cm is the average flight distance from nozzle to skimmer, and $n_{bg}=p_{bg}/k_B T$ and $v_{bg}=\sqrt{8k_B T_0/(\pi m)}$ are the residual gas density and mean thermal velocity, respectively. The Kr-Kr collision cross section $\sigma$ is varied as a free fit parameter, yielding an average value $\sigma = 130\pm 30\mathrm{\AA}^2$. This value is in reasonable agreement with the Landau-Lifshitz approximation for the maximum scattering cross section, $\sigma_{max}=8.083\left(C^{(6)}/(\hbar v_{rel})\right)^{2/5}=390\mathrm{\AA}^2$, with $C^{(6)}=77.4\,$eV$\mathrm{\AA}^6$ for Kr~\cite{Levine:2005}.

\begin{figure}
\includegraphics[width=9cm]{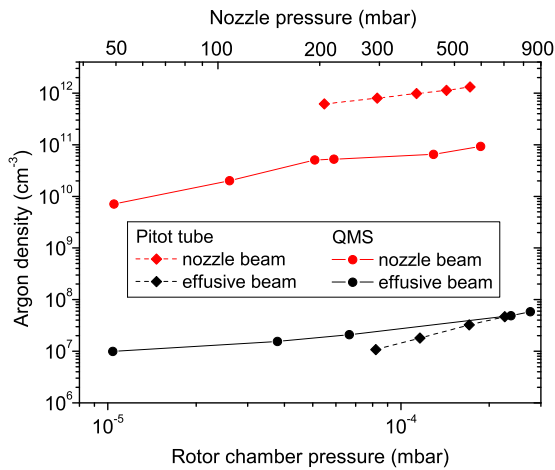}
\caption{\label{fig:effusive}Comparison of beam densities of the continuous nozzle beam (rotor at rest) and the effusive beam produced by the residual gas in the rotor chamber. Circles depict the data measured using the quadrupole mass spectrometer scaled to densities obtained from the detector chamber pressure measured using a cold cathode pressure gauge. Squares represent the density values obtained from the Pitot tube measurement.
}
\end{figure}
High nozzle pressures, which lead to increased rest-gas pressure in the nozzle chamber, cause another unwanted effect to occur -- the formation of a continuous effusive beam. The latter adds up to the pulsed nozzle beam and may have comparable beam intensities as the nozzle beam is decelerated to low speeds and low densities. We therefore compare the densities of nozzle beams and effusive beams by recording the detector signal with the nozzle standing properly aligned in front of the skimmer, on the one hand, and with the rotor being tilted completely out of the beam axis, on the other. The resulting density values are plotted in Fig.~\ref{fig:effusive}. While the measurement using the Pitot tube yields higher nozzle beam densities by a factor of 10, equal densities are found for the effusive beam. Thus, the nozzle beam is higher in density by 3-4 orders of magnitude than the effusive beam.

For decelerated beams of Ar and Kr, however, equal densities of nozzle and effusive beams are reached at beam velocities $v_0\lesssim 60$ and $v_0\lesssim 40$\,m/s, respectively. Provided the expansion is operated in the supersonic regime, the ratio of nozzle and effusive beam densities is independent of the nozzle pressure $p_0$. This ratio can be improved either by increasing the pumping speed in the nozzle chamber, which is characteristic for such an arrangement based on a continuous gas expansion. The effusive background can be further suppressed by implementing a mechanical velocity selector synchronized with the rotating nozzle. Therefore, in the near future a chopper wheel will be placed behind the skimmer.

\section{\label{sec:Guide}Electrostatic guiding}
\begin{figure}
\includegraphics[width=9cm]{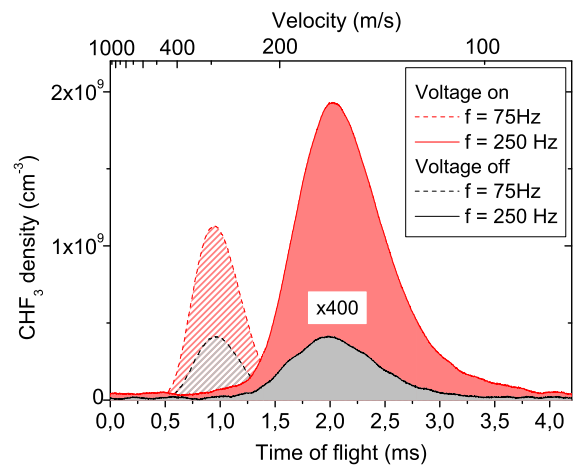}
\caption{\label{fig:EnhancementTOF}Typical CHF$_3$ density distributions as a function of flight time for two different speeds of rotation of the nozzle. The data recorded at 250\,Hz are scaled up by a factor of 400. Signals shown as back lines are recorded with no voltage applied to the guide electrodes, light red lines are obtained at 3.5\,kV electrode voltage.}
\end{figure}
The issue of transverse divergence of decelerated beams can be relieved for beams of polar molecules by implementing additional guiding elements using inhomogeneous electrostatic fields. This technique has been used in the past for focussing and even state selecting molecular beams for reactions dynamics experiments~\cite{Bernstein:1982}. More recently, the technique has been used for guiding and trapping of cold polar molecules produced by Stark deceleration or Stark filtering~\cite{Bethlem1:1999,Rangwala:2003,Junglen:2004,Krems:2009}.

In our setup we have implemented an electrostatic quadrupole guide between the skimmer and the ionizer of the QMS. It consists of four 259\,mm long stainless steel rods 2\,mm in diameter with a gap of 2\,mm between two diagonally opposing rods. In a first experiment, slow beams of fluoroform (CHF$_3$) molecules have been produced by expanding CHF$_3$ either as a neat gas or seeded in Kr out of the rotating nozzle. CHF$_3$ is a polar molecule with a permanent dipole moment of 1.65\,Debye. Since its mass ($m_{CHF3}\approx 70\,$amu) lies close to the mass of Kr ($m_{Kr}\approx 80\,$amu) it can be efficiently cooled and decelerated by coexpansion with Kr.

When applying a high voltage $U=3.5\,$kV to a pair of opposing electrodes the QMS detector signal clearly rises, as depicted in Fig.~\ref{fig:EnhancementTOF}. In these measurements, CHF$_3$ is expanded at a nozzle pressure $p_0=200\,$mbar at rotor frequencies 75 and 250\,Hz, leading to most probable beam velocities $v_w=450$ and $v_w=165\,$m/s, respectively. Clearly, the enhancement of the QMS signal at an electrode voltage $U=3.5\,$kV in proportion to the signal with no high voltage is higher at low beam velocities, reaching roughly a factor of 5 for $v_w=165\,$m/s. Great care was taken to suppress any direct influence of the electrode voltage on the ionization efficiency of the QMS ionizer.

Similar experiments have also been performed with deuterated ammonia (ND$_3$). In addition to increasing the beam intensity, the quadrupole fields act as quantum state selectors which filter out those states that are attracted towards the beam axis where the electric field amplitude is minimal (low-field seeking states). Provided the rotational temperature of the ND$_3$ molecules is comparable to the translational one, \textit{i.\,e.} in the range of a few Kelvin, we may expect to produce ND$_3$ molecules in the state $J=1$, $M=K=-1$ with very high purity~\cite{Bethlem:2002}.

\begin{figure}
\includegraphics[width=9cm]{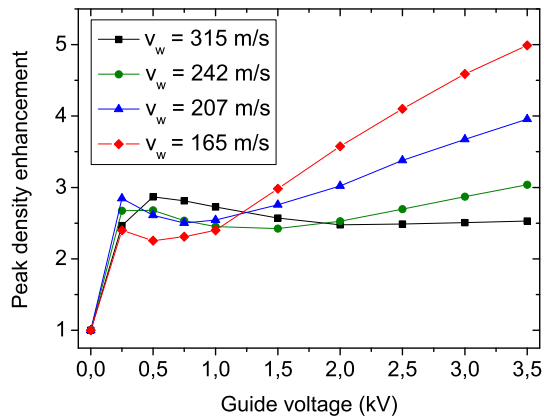}
\caption{\label{fig:EnhancementVoltage}
Relative efficiency of guiding CHF$_3$ molecules at various beam velocities as a function of electrode voltage.}
\end{figure}
Interestingly, the dependence of the enhancement of peak density on the electrode voltage $U$ features a non-monotonic increase, as shown in Fig.~\ref{fig:EnhancementVoltage}. A first local signal maximum is observed at electrode voltages between 200 and 500\,V, depending on beam velocity. We interpret this behavior in terms of two competing mechanisms being active: Guiding of molecules in low-field seeking states enhances the detector signal whereas deviation of high-field seeking molecules away from the beam axis reduces the measured beam density. At low voltages $U\approx 500\,$V, the guiding effect outreaches the deviation efficiency, leading to a local density maximum. As the electrode voltage rises, all high-field seeking molecules are expelled out of the beam and only low-field seekers are guided to the detector. This guiding effect partly compensates the transverse blowing up of the beam, which is most relevant for low longitudinal beam velocities. Consequently, the relative guiding efficiency increases up to a factor of 5 as the beam velocity is reduced down to $v_w=165\,$m/s. The fact that the same qualitative behavior is observed with beams of ND$_3$ supports our interpretation, which does not consider specific Stark shifts of the molecules. In future efforts the transmission of our setup will be further studied both experimentally and using molecular trajectory simulations. Besides, guiding of decelerated molecules around bent electrodes will be implemented to provide state-purified beams of cold molecules. This has the additional advantage of efficiently suppressing the hot effusive background.

\begin{figure}
\includegraphics[width=9cm]{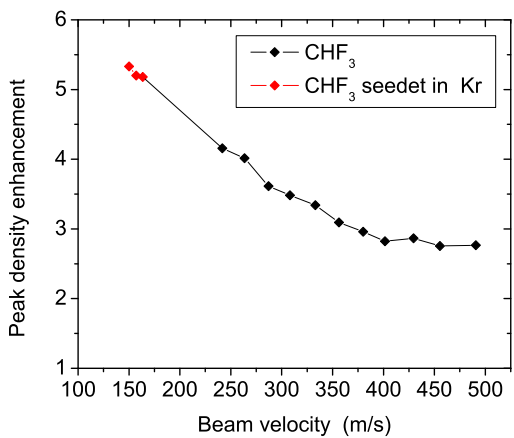}
\caption{\label{fig:EnhancementRot}Enhancement of the CHF$_3$ peak density as a function of beam velocity for fixed electrode voltage $U=3.5$\,kV.}
\end{figure}
The dependence of the enhancement factor on the beam velocity is illustrated in Fig.~\ref{fig:EnhancementRot}. In this measurement, the electrode voltage is set to the maximum value $U=3.5\,$kV and the beam velocity is tuned from $v_w=500$ down to $v_w=250\,$m/s by changing the frequency of rotation of the nozzle. Further deceleration down to $v_w\approx 150\,$m/s is obtained by seeded expansion of CHF$_3$ in Kr at a mixing ratio 1:2. The enhancement factor continuously rises as the beam velocity is reduced, which highlights the potential of combining decelerated nozzle beams with electrostatic guiding fields.

\section{\label{sec:Conclusion}Conclusion and outlook}
In conclusion, we have presented a versatile apparatus that produces cold beams of accelerated or decelerated molecules based on a rapidly rotating nozzle. Various technical improvements with respect to the original demonstration by Gupta and Herschbach~\cite{Gupta:1999,Gupta:2001} are introduced. In particular, gas injection into the rotor is now realized using a fluid metal-sealed rotary feedthrough which eliminates gas leakage into the vacuum chamber. Using this setup, beam velocities well below $100$\,m/s and longitudinal beam temperatures down to about 1\,K are achieved. A detailed analytic model to simulate the measured density distributions as functions of flight time is presented. The fundamental drawback of this technique, the sharp drop of beam intensity at slow beam velocities, is relieved by combining the setup with electrostatic guiding elements, provided the molecules feature a suitable Stark effect. Moreover, Stark guiding goes along with internal-state selectivity. Thus, a significant increase of beam intensity as well as an expected high state-purity is achieved with beams of fluoroform and ammonia molecules.

Thanks to the technical improvements the rotating nozzle setup has evolved into a simple and reliable source of cold molecular beams with tunable velocity. In particular, slow beams of molecules that are not amenable to efficient filtering or deceleration by means of the Stark interaction with external electric fields can now be produced with high intensity. In the near future, we will utilize this beam source for studying reactive collisions between slow molecules and cold atoms produced by other means, \textit{e.\,g.} using a magneto-optical trap. At collision energies down to $\lesssim 1\,$meV, interesting new quantum effects may be expected in the reaction dynamics, \textit{e.\,g.} strong modulations of the scattering cross-section in reactions of the type Li+HF$\rightarrow$LiF+H~\cite{Weck:2005}. Furthermore, intense cold molecular beams with tunable velocity and internal-state purity are of current interest for surface scattering experiments~\cite{Liu:2004}.

\begin{acknowledgments}
We thank H.J. Loesch for ceding to us the main part of the experimental setup. Generous advice by D. Weber, K. Zrost and M. DeKieviet as well as assistance in setting up the apparatus by J. Humburg, Ch. Gl\"uck and O. B\"unermann is gratefully acknowledged. We are grateful for support by the Landesstiftung Baden-W\"urttemberg as well as by DFG.
\end{acknowledgments}

\newpage 

\end{document}